\documentclass[aps,epsfig,manuscript,showpacs]{revtex4}
\usepackage{amsmath,amssymb,graphicx}
\usepackage{epsfig}
\usepackage{graphicx}
\usepackage{enumerate}
\newcommand{\beq}{\begin{equation}}
\newcommand{\eeq}{\end{equation}}

\newcommand{\bea}{\begin{eqnarray}}
\newcommand{\eea}{\end{eqnarray}}

\begin{document}

\title{Molecular heat pump}

\author{Dvira Segal$^1$ and Abraham Nitzan$^2$. }

\affiliation{$^1$Department of Chemical Physics, Weizmann
Institute of Science, 76100 Rehovot, Israel \\
$^2$School of Chemistry, Tel Aviv University, Tel Aviv, 69978,
Israel}

\begin{abstract}
We propose a novel molecular device that pumps heat against a
thermal gradient. The system consists of a molecular element
connecting two thermal reservoirs that are characterized by
different spectral properties. The pumping action is achieved by
applying an external force that periodically modulates molecular
levels. This modulation affects periodic oscillations of the
internal temperature of the molecule and the strength of its
coupling to each reservoir resulting in a net heat flow in the
desired direction. The heat flow is examined in the slow and fast
modulation limits and for different modulation waveforms, thus
making it possible to optimize the device performance.
\end{abstract}

\pacs{63.22.+m, 44.10.+i, 05.60.-k, 66.70.+f }

\date{\today}
\maketitle

\section{Introduction}
A heat pump is a device that transfers heat from a low to a high
temperature reservoir by applying an external work that modulates
the system's parameters. This paper discusses a molecular machine
of this kind. The analogous electrical device that transfers
charge (or spin) against the electrochemical potential bias was
studied theoretically \cite{Thouless} and demonstrated
experimentally in an open quantum dot when varying both the dot
voltage and the tunneling barrier heights \cite{Switkes}.

In a prototype particle pumping machine, each cycle begins with
isolating the system from one reservoir by reducing its coupling
to the system, while applying a potential that drives carriers
from the other reservoir into the system. Next, this configuration
is reversed, the system is coupled to the previously disconnected
reservoir and isolated from the previously connected one, and its
potential changes so as to drive carriers from the system into the
connected reservoir. Consequently a net current is flowing through
the system. A basic requirement for demonstrating pumping
operation is the modulation of at least {\it two} internal
parameters. Out of phase modulation provides an adiabatic
(reversible) pumping operation \cite{Altshuler}, while in the
general case, quasiadiabatic (irreversible) processes can be
realized \cite{Astumian}.

Motivated by the growing interest in nanomechanics \cite{Blencowe}
and quantum thermodynamics \cite{Nieu}, we present here a
molecular model for a {\it thermal} pump that is based on similar
operating principles. Other thermal devices that have been
envisioned recently are a heat rectifier \cite{Casati1, Casati2,
Spinboson}, a thermal transistor, \cite{Casati3} and even a
mechanical analog of a laser \cite{Bargatin}. As with any machine,
one seeks optimization of performance with respect to both
efficiency and power.

In our model a molecular unit connects two spatially separated
left ($L$) and right ($R$) heat baths held at different
temperatures, and transfers heat from the cold ($C$) (henceforth
referred to as the left side) into the hot ($H$) (right side)
reservoir. An external force modulates the energy level structure
of the conducting molecule and consequently its effective coupling
to the reservoirs (thus providing a modulation of two system
parameters while modulating a single physical variable). This
system is shown to operate as a heat pump that can transfer energy
from a cold to a hot reservoir.

Similar abstract models of this nature were proposed before by
Kosloff and coworkers \cite{Kosloff84, Geva, Pablo, Feldmann}.
Here we consider a specific, realizable, model of a molecular
level heat pump based on modulation of molecular energy levels.
Such modulation can be achieved by a stark shift affected by a tip
induced local electric field, by magnetic field splitting of
energy levels and by an external force applied by the tip of an
atomic force microscope \cite{AFM}. It was also demonstrated
recently that nanotubes tension can be tuned by applying an
electric field, thus modulating the tube vibrational frequencies
\cite{Yaish, Roundy}. Finally, compression of molecules affects
their vibrational modes, e.g. the radial breathing modes of
nanotubes are pressure dependent \cite{Strain} with about
$d\omega/ dP \sim 1$ cm$^{-1}$/Gpa \cite{RBM}, making high
pressures necessary for a significant effect. Each of these
schemes can be used as a basis of the proposed heat engine. Below
we describe the concept of this engine, consider its performance
and efficiency in terms of molecular and junction parameters, and
suggest possible optimization methods.

\section{Model}

The model system consists of a molecular unit connecting two
thermal reservoirs $L$ and $R$ of inverse temperatures
$\beta_L=(k_BT_L)^{-1}$ and $\beta_R=(k_BT_R)^{-1}$ respectively,
where $k_B$ is the Boltzmann constant. For simplicity we assume
that heat transfer is dominated by a specific single mode. In
addition, if the baths temperatures are low enough, only the
lowest vibrational states of the molecule are populated, and we
can model the isolated molecule by a two level system (TLS). An
external force drives periodically the frequency of this molecular
mode, i.e. the two level energy spacing. The total Hamiltonian
therefore includes three terms
\beq H=H_S+H_B+H_{MB},
\label{eq:Hamilt} \eeq
where
\beq H_S=\frac{\omega(t)}{2}\left(|1 \rangle \langle 1| -|0
\rangle \langle 0| \right) \label{eq:Hs} \eeq
is the Hamiltonian of the molecular mode under consideration
($\hbar\equiv 1$). Here $|0 \rangle$ and $|1\rangle$ represent the
two  states of energies $\epsilon_0$ and $\epsilon_1$, and
\beq \omega(t)\equiv \epsilon_1-\epsilon_0=\omega_0+ F(t),
\label{eq:omt}
\eeq
provides the time dependent driving with a static frequency
$\omega_0$ and a periodic modulation
$F(t)=F(t+\frac{2\pi}{\Omega})$. In what follows we refer to
$\omega(t)$ as the instantaneous energy gap. $F(t)$ can be
expanded in a Fourier series
\beq
F(t)=\sum_{n=-\infty}^{\infty}\left[A_n\cos(n\Omega t) +C_n\sin(n\Omega t) \right].
\label{eq:Fom}
\eeq
We also define the indefinite integral of this perturbation that will be useful
below
\bea
f(t)&\equiv&\int F(t)dt
\nonumber\\
&=&\sum_{n=-\infty}^{\infty}\left[\frac{A_n}{n\Omega}\sin(n\Omega
t)- \frac{C_n}{n\Omega}\cos(n\Omega t) \right].
\label{eq:Fomt}
\eea
The two thermal reservoirs $L$ and $R$
%
\bea
H_B&=&H_L+H_R
\label{eq:HB}
\eea
%
do not interact directly with
each other, and can exchange energy only through their coupling to
the system. Transitions between the $|0 \rangle \leftrightarrow
|1\rangle$ states can occur due to the coupling to these heat
baths
\bea
H_{MB}&=&B|0\rangle \langle1| +B^{\dagger}|1\rangle \langle0|,
\nonumber\\
B&=&B_L+B_R,
\label{eq:HMB}
\eea
where $B_K$ ($K=L,R$), the bath operators, are given in terms of
their phonon coordinates. The thermal reservoirs are characterized
by their spectral density functions. An essential ingredient of
our model is having different spectral functions for the left and
right reservoirs. Below we model this difference by assuming that
the reservoirs are characterized by different Debye frequencies
$\omega_D^L \neq \omega_D^R$. Similar effects may be achieved by
other means, e.g., connecting identical bathes to the system via
'doorway oscillators' of different frequencies.

Eqs.~(\ref{eq:Hamilt})-(\ref{eq:HMB}) represent a particular kind
of a molecular relaxation process \cite{VER}.  Unlike the standard
relaxation models here the molecular  mode is (a) coupled to {\it
two} thermal reservoirs of different temperatures and spectral
properties, and (b) modulated by an external force so that the
corresponding level spacing  oscillates in time. This then becomes
a "driven dissipative" system that differs from previously
considered models \cite{Milena1} by working in the system
eigenstate representation and by coupling to {\it two} independent
thermal reservoirs.

\section{Operation cycle}

\begin{figure}[htbp]
\vspace{0mm} \hspace{1mm}
{\hbox{
\vspace*{3mm}
\epsfxsize=80mm \epsffile{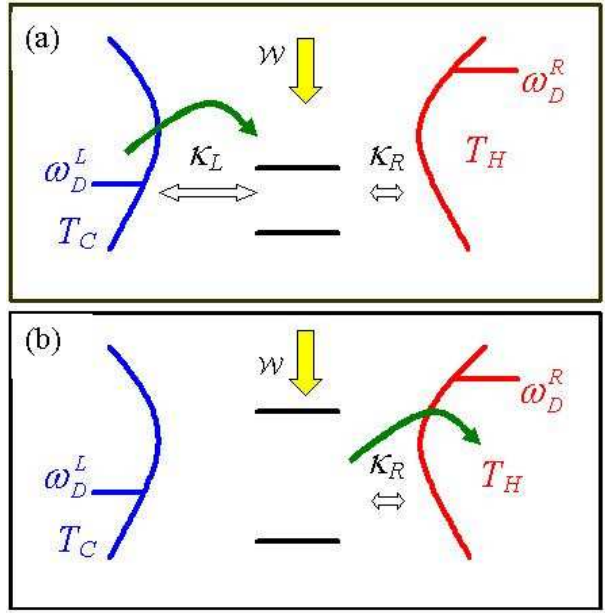}}}
\caption{(Color online) Schematic picture of the pumping cycle.
(a) At low frequencies the TLS is strongly coupled to the left reservoir. Thus when
$T_{TLS}<T_L$ heat is transferred from the $L$ bath to the TLS.
(b) At high frequencies the TLS is coupled only to the right reservoir, thus its
internal energy is transmitted into the right bath when $T_{TLS}>T_R$. }
\label{Fig0}
\end{figure}

Next we describe a setup that leads to the desired pumping
operation. First, a temperature gradient is applied  across the
system by keeping $T_R \equiv T_H>T_L \equiv T_C$. ($H$ and $C$
stand for "hot" and "cold" reservoirs). In addition, asymmetry is
built into the system by choosing $\omega_D^L<\omega_D^R$ and
 $\kappa_L>\kappa_R$, where $\kappa_K$ is a parameter related to
 the vibrational relaxation rate induced by the $K$ thermal bath (see Eqs.
 (\ref{eq:expnad}) and (\ref{eq:expad})).
With this choice of parameters, when the TLS frequency $\omega(t)$
is small, it is coupled more strongly to the left, cold reservoir.
Energy is then injected from the left reservoir into the system
whenever the TLS temperature defined as
\beq
T_{TLS}(t)=-\frac{\omega(t)}{k_B {\rm log}(P_1(t)/P_0(t))}
\label{eq:TTLS}
\eeq
is smaller than $T_C$. Here $P_{0}$ and $P_1$ are the population
of the $|0\rangle$ and $|1\rangle$ states respectively. The TLS
energy spacing is next increased by the action of the external
force, therefore it couples more effectively to the right, hot
reservoir. If the levels population is kept (almost) fixed during
this process, the effective TLS temperature becomes very high. If
it is higher than $T_R \equiv T_H$, heat will be transferred from
the TLS into the right -hot reservoir- and the pumping cycle is
completed. For a schematic representation see Fig.~\ref{Fig0}.

This pumping machine is a continuous version of the discrete four
strokes pump of Ref.~\cite{Feldmann}. Here the system is
effectively disconnected from each reservoirs at different times
due to the asymmetric construction of the reservoirs spectral
properties and the system-bath interactions.


\section{Dynamics}

Given the time dependent Hamiltonian,
Eqs.~(\ref{eq:Hamilt})-(\ref{eq:HMB}), a Master equation for the
states population $P_n$ ($n$=0,1) can be obtained by making the
following assumptions: (i) The system-heat bath couplings are
small so that second order perturbation theory can be applied to
yield golden-rule-type relaxation rates. (ii)
The memory time of the bath fluctuations $\tau_K$ ($K$=$L$,$R$) is
short relative to the thermal relaxation time
\bea \tau_K^{-1}\gg \Gamma.
\label{eq:approx} \eea
Here $1/\Gamma$ is the thermal relaxation time of the two-level
system given by $\Gamma^{-1}=\left(k_u+k_d\right)^{-1}$, see definitions
below. Under these assumptions, Redfield theory \cite{Redfield}
leads to the Markov-Master equations for the states population,
(See Appendix A for a detailed derivation)
\begin{eqnarray}
 \dot{P_1}&=&-  k_{d}(t) P_1+k_{u}(t) P_0;\,\,\
\nonumber\\
P_1&+&P_0=1,
\label{eq:Master}
\end{eqnarray}
where
\bea k_u&=&\int_{0}^{t} e^{i\omega_0(t-\tau)}
e^{i[f(t)-f(\tau)]} \langle B(\tau)B^{\dagger}(t)\rangle d\tau
\nonumber\\
&+& \int_{0}^{t} e^{-i\omega_0(t-\tau)}
e^{-i[f(t)- f(\tau)]}
\langle B(t)B^{\dagger}(\tau)\rangle d\tau,
\label{eq:kup}
\eea
\bea
 k_d&=&\int_{0}^{t} e^{i\omega_0(t-\tau)} e^{i[ f(t) -f(
\tau)]} \langle B^{\dagger}(t)B(\tau) \rangle d\tau
\nonumber\\
&+& \int_{0}^{t} e^{-i\omega_0(t-\tau)} e^{-i[f(t)-f(\tau)]}
\langle B^{\dagger}(\tau)B(t)\rangle d\tau.
\label{eq:kdown}
\eea
Here $f(t)$ is the time periodic function of Eq.~(\ref{eq:Fomt}).
Note that the relaxation rates include contributions from both
left and right thermal baths since $\langle B^{\dagger}(t)
B(0)\rangle= \langle B_L^{\dagger}(t)B_L(0)\rangle_L + \langle
B_R^{\dagger}(t) B_R(0)\rangle_R$, as implied by
Eq.~(\ref{eq:HMB}) where the averages are over the thermal
distributions of the corresponding baths. Therefore
\bea
k_u=k_{u,L}+k_{u,R}; \,\,\,
k_d=k_{d,L}+k_{d,R}.
\label{eq:ratesum}
\eea
%

\subsection{Rate constants: General expression}

We derive next explicit expressions for the rate constants in the
general, {\it non adiabatic} regime. We begin with the first
integral of the excitation rate $k_u$ in Eq.~(\ref{eq:kup}). It
can be expanded as follows
\bea I_1\equiv\sum_{n,m} J_mJ_n^* e^{i(m-n)\Omega t}
\int_{0}^{\infty}e^{i(\omega_0+n\Omega)x} \langle B(0)
B^{\dagger}(x)\rangle dx,
\nonumber\\
\label{eq:I1}
\eea
where the sum goes over $\sum_{n,m= -\infty}^ {\infty}$, and
the upper limit in the integral is extended to infinity. To obtain
(\ref{eq:I1}) we have utilized the Fourier expansion
\bea
 e^{if(t)}&=&\sum_{n=-\infty}^{\infty}J_ne^{ i n\Omega t}.
\eea
For the simple cosine modulation, $F(t)=A_1 \cos(\Omega t)$, the
expansion coefficients reduce to the Bessel functions
$J_n(A_1/\Omega)$ of order $n$. We note that the $m$ series in
Eq.~(\ref{eq:I1}) trivially sums up to $e^{if(t)}$, yet we prefer
this representation since it formally eliminates the specific
$F(t)$ dependence from the equations.  The second integral in Eq.
(\ref{eq:kup}) can be manipulated in the same way to produce
\bea I_2&\equiv& \sum_{n,m} J_n J_m^* e^{i(n-m)\Omega t}
\int_{-\infty}^{0}e^{i(\omega_0+n\Omega)x} \langle B(0)
B^{\dagger}(x)\rangle dx.
\nonumber\\
\label{eq:I2}
\eea
Assuming that $\int_{-\infty}^{\infty}e^{i(\omega_0+n\Omega)x}
\langle B(0) B^{\dagger}(x)\rangle dx$ is a symmetric function
around zero so that the integrals in $I_1$ and $I_2$ can be
replaced by $\frac{1}{2}\int_{-\infty}^{\infty}...$, the total
excitation rate becomes
\bea
 k_u&\equiv& I_1+I_2
= \sum_{n,m} \Re\left[J_n J_m^*
e^{  i(n-m)\Omega t}  \right]  k_{u}^{(n)};
\nonumber\\
 k_{u}^{(n)}&=&\int_{-\infty}^{\infty}e^{i(\omega_0+n\Omega)x}
\langle B(0) B^{\dagger}(x)\rangle dx,
\label{eq:ku}
 \eea
where $\Re$ denotes the real part.
 It is given in terms of  standard time independent transition
rates evaluated at different overtone frequencies $n\omega$ ($n$
are integers), multiplied by the appropriate Fourier coefficients
and a time dependent modulation.
 The downward
rate $k_d$ is obtained in a similar way by combining the two
integrals of Eq.~(\ref{eq:kdown})
\bea
k_d &=&\sum_{n,m} \Re \left[J_n J_m^*
 e^{i(n-m)\Omega t} \right] k_{d}^{(n)};
\nonumber\\
 k_{d}^{(n)}&=& \int_{-\infty}^{\infty}e^{i(\omega_0+n\Omega)x}
\langle B^{\dagger}(x)B(0)\rangle dx.
\label{eq:kd}
\eea
For infinitely slow modulation, $A_n/\Omega, C_n/\Omega  \rightarrow \infty$,
 the standard time independent expression for the vibrational relaxation (VR)
  rate \cite{Skinner} is recovered, $k_d(t) \rightarrow  k_d^{(0)}$, by making use of the sum
identity $\sum_{n,m} J_nJ_m^*e^{i(n-m)\Omega t}=1$.

The transition rates (\ref{eq:ku}) and (\ref{eq:kd}) can be
further decomposed into the $L$ and $R$ contributions as in Eq.
(\ref{eq:ratesum}). The up and down rates induced by each thermal
reservoir are interrelated by the detailed balance condition for
each $n$ component
\beq
k_{u,K}^{(n)}= k_{d,K}^{(n)} e^{-\beta_K (\omega_0+n\Omega)}, \,\,\,\, (K=L,R).
\label{eq:detail}
\eeq

Our results so far are general within the weak system-reservoir
interaction limit. As a specific model for the bath correlation
functions we invoke below the exponential energy gap form
\cite{VERNitzan}
\bea
k_{d,K}^{(n)}=
\begin{cases}
\kappa_{K}, &  \omega_0+n\Omega< \omega_D^K \\
\kappa_{K} e^{-(\omega_0+n\Omega)/\omega_D^K}, &  \omega_0+n\Omega>\omega_D^K.\\
\end{cases}
\label{eq:expnad}
\eea
Here $\omega_D^K$ is the Debye frequency characterizing the
$K=L,R$ reservoir. In general we assume $\kappa_L \neq \kappa_R$
and $\omega_D^L\neq \omega_D^R$. We associate the bath relaxation
time $\tau_K$ with the inverse Debye frequency.
\subsection{Rate constants: Adiabatic regime}

We can also derive explicit expressions for the transition rates
assuming the energy modulation is adiabatic, i.e. does not itself
induce transitions in the TLS or in the thermal reservoirs.  In
this regime the integrals of Eqs. (\ref{eq:kup})-(\ref{eq:kdown})
can be simplified
by approximating the differences by first derivatives,
$f(t)-f(t-x) \sim xF(t)$, $x=t-\tau$. Higher order terms are neglected
assuming $|\frac{1}{F(t)}\frac{dF(t)}{dt}|\ll \omega_D^K$.
(For a cosine modulation, $F(t)=A_1\cos(\Omega t)$, this condition
translates into $\Omega\ll \omega_D^K$).
Then
\bea I_1&=&\int_{0}^{\infty} e^{i\omega_0 x} e^{i \left[f(t)-
f(t-x)\right]}
\langle B(0)B^{\dagger}(x)\rangle dx
\nonumber\\
&
\longrightarrow& \int_{0}^{\infty} e^{i\omega_0 x}
e^{ixF(t)} \langle B(0)B^{\dagger}(x)\rangle dx.
\label{eq:AdiabaticI1} \eea
Conducting similar operations on the second integral of Eq.~(\ref{eq:kup}) yields
\beq I_2
 =\int_{-\infty}^{0} e^{i\omega_0 x} e^{ixF(t)}
\langle B(0)B^{\dagger}(x)\rangle dx,
\label{eq:AdiabaticI2}
\eeq
and the adiabatic rate constants become
\bea
 k_{d,K}&=& \int_{-\infty}^{\infty} d\tau e^{i \omega(t) \tau}
\langle B^{\dagger}_{K}(\tau) B_{K}(0) \rangle,
\nonumber\\
k_{u,K}&=&k_{d,K} e^{-\beta_K \omega(t)}; \,\,\, \omega(t)=\omega_0+F(t).
\label{eq:rateAd}
\eea
The adiabatic approximation therefore implies that an
instantaneous detailed balance would be satisfied at all times if
the system was coupled to a single bath, i.e.
$P_1(t)/P_0(t)=e^{-\omega(t)/k_BT}$, where $T$ is the single bath
temperature. Also in this case we utilize the exponential energy
gap law for modeling the adiabatic relaxation rates
\bea k_{d,K}=\begin{cases}
\kappa_K, &  \omega(t)< \omega_D^K \\
\kappa_K e^{-\omega(t)/\omega_D^K}, &  \omega(t)>\omega_D^K.\\
\end{cases}
\label{eq:expad}
\eea
%

\section{Analysis of performance}

The performance of a heat pump can be characterized both in terms of its power-
the amount of heat transferred per cycle (i.e. period of the modulating force)
and its efficiency, defined as the ratio between the heat transferred and the
work invested. The internal energy $\mathcal{E}$ of the TLS is given by
\bea
\mathcal{E}= \epsilon_0 P_0+\epsilon_1 P_1,
\eea
where $\epsilon_0$ and $\epsilon_1$ are measured from some fixed
reference, here chosen by Eq.~(\ref{eq:Hs}) to be the midpoint
between the two levels. The internal energy is changed either
through modulation of the TLS energy spacing or due to population
transfer between the levels \cite{Kosloff84, Feldmann}
\bea
\frac{d\mathcal{E}}{dt}=\epsilon_0\frac{dP_0}{dt}  +\epsilon_1\frac{dP_1}{dt}
+P_0\frac{d\epsilon_0}{dt}+ P_1\frac{d\epsilon_1}{dt}.
\label{eq:dedt}
\eea
This rate of energy change can be separated into its work $ \dot
{\mathcal{W}}$, and heat $ \dot {\mathcal{Q}}$ components
\bea
\dot{\mathcal{W}}\equiv P_0 \frac{d \epsilon_0}{dt}+ P_1 \frac{d
\epsilon_1}{dt},
\nonumber\\
\dot{\mathcal{Q}}\equiv \epsilon_0\frac{dP_0}{dt} +\epsilon_1\frac{dP_1}{dt}.
\label{eq:QW}
\eea
Using the following equalities
\bea
\dot{P_0}+\dot{P_1}=0; \,\,\, \dot{\epsilon_0}+\dot{\epsilon_1}=0,
\eea
that are based on Eqs.~(\ref{eq:Hs}) and (\ref{eq:Master}), we
find
\bea
\dot{\mathcal{W}}= S(t) \dot{\omega}(t); \,\,\,\,
\dot{\mathcal{Q}}= \omega(t) \dot{S}(t),
\label{eq:QWSymm}
\eea
where $S\equiv (P_1-P_0)/2$ is referred to as the system
polarization \cite{Kosloff84, Feldmann}. As the effect of the two
reservoirs is additive, we can decompose the rate at which $S$
changes to its $L$ and $R$ contributions
\bea
\dot{S}&=&\dot S_{L}+\dot S_{R},
\nonumber\\
\dot S_{K}&=& -k_{d,K}P_1 + k_{u,K}P_0, \,\,\, (K=L,R),
\label{eq:P1LR}
\eea
where $P_1$ and $P_0$ are obtained by solving Eq. (\ref{eq:Master}).
Consequently, the heat flux $\dot{\mathcal{Q}}$ can be written as
a sum of $L$ and $R$ terms
\beq
\dot{\mathcal{Q}} \equiv \dot{\mathcal{Q}}_{L}+
\dot{\mathcal{Q}}_{R}; \,\,\ \dot{\mathcal{Q}}_{K}=\omega(t)\dot
S_{K}.
\label{eq:QLR}
\eeq
We note that in steady state
$\dot{\mathcal{Q}}_{L}=-\dot{\mathcal{Q}}_{R}$, i.e. the heat
current is the same at the left and right contacts. Here these
quantities are in general different, even on the average, due to
the action of the external perturbation, $J_L=\dot{\mathcal{Q}}_L
\neq J_R=-\dot{\mathcal{Q}}_R$. In the equations above the heat
current is taken positive when flowing left to right.

The coefficient of performance (COP) of a  heat transfer machine
can be defined with regard to its performance either as a heat pump
\beq \eta_{H}=\mathcal{Q}_{H}/\mathcal{W}, \eeq
or as a refrigerator
\beq
\eta_{C}=\mathcal{Q}_{C}/\mathcal{W},
\label{eq:etaC}
\eeq
where $\mathcal{Q}_K=\int_{cycle}\dot{\mathcal{Q}}_K$ ($K=C,H$) and
$\mathcal{W}=\int_{cycle}\dot{\mathcal{W}}$.
The maximal theoretical values of these coefficients are given by that of
a reversible (Carnot) machine,
\beq
\eta^{max}_{H}=
\frac{\mathcal{Q}_{H}}{\mathcal{Q}_{H}-\mathcal{Q}_{C}} =
\frac{T_{H}}{T_{H}-T_{C}},
\eeq
\beq
\eta^{max}_{C}=
\frac{\mathcal{Q}_{C}}{\mathcal{Q}_{H}-\mathcal{Q}_{C}} =
\frac{T_{C}}{T_{H}-T_{C}}.
\eeq
In what follows we focus on the refrigerator COP, Eq.
(\ref{eq:etaC}), as a measure of efficiency of our molecular
machine.

In an {\it ideal} refrigerator the operation cycle
consists four distinct steps: (i)
Thermal: The TLS with an energy gap $\omega_L$ couples to, and
exchanges energy with, the left (cold) bath only. (ii) Adiabatic:
The TLS is decoupled from the reservoirs and its energy spacing is
increased to $\omega_R$. (iii) Thermal: The TLS is coupled to, and
exchanges energy with, the right (hot) reservoir only. (iv)
Adiabatic: The TLS, again decoupled from the reservoirs, restores
its energy gap back to the low $\omega_L$ value.
It should be emphasized that in the realizable
machine discussed in Section III, system bath decoupling is not imposed.
It is  approximated by replacing the
adiabatic steps by transitions whose durations are short relative to the thermal
relaxation time associated with the system-bath coupling.
Optimized performance is therefore obtained when the adiabatic
branches of the process are  fast relative to the thermal branches, so that no
backward heat flow takes place. In contrast, the thermal branches
should be long enough for attaining full equilibration of the TLS
with the interacting bath.

Consider first the ideal refrigerator in which the system is decoupled
from the left/right reservoirs during the adiabatic branches. In
this limit Eq.~(\ref{eq:Master}) can
be solved analytically for the thermal branches yielding the time
evolution of the $K=L,R$ polarization \cite{Feldmann}
\beq S_K(t)=S_K^{eq}+\left( S_K(0)-S_K^{eq} \right)e^{-\Gamma_Kt}; \,\,\ K=L,R,
\label{eq:SK}
\eeq
where $S_K^{eq}$ is the equilibrium polarization
\beq
S_K^{eq}=-\frac{1}{2}\tanh (\omega_K/2k_BT_K).
\label{eq:Seq}
\eeq
$\omega_K$ is the time independent TLS gap when it is connected to
the $K$ reservoir and $\Gamma_K=k_{d,K}+k_{u,K}$.
We denote the durations of the thermal branches,
 i.e. the contact times of the system with the $L$ and $R$ reservoirs
by $\tilde \tau_L$ and $\tilde \tau_R$ respectively,
 and recall that the polarization $S=(P_1-P_0)/2$ does not change during
 the (ideal) adiabatic branches.
 The polarizations at the beginning of the thermal branches are therefore
given by $S_L(0)=S_R(\tilde\tau_R)$ and  $S_R(0)=S_L(\tilde
\tau_L)$.
Using these initial values  in
Eq.~(\ref{eq:SK}), the solution of the two coupled linear equations at
 the end of the thermal branches, i.e. at $t=\tilde \tau_K$ is
\beq S_R(\tilde \tau_R)=
S_L^{eq}+\frac{(S_R^{eq}-S_L^{eq})(1-e^{-\Gamma_R\tilde
\tau_R})}{1-e^{-\Gamma_L\tilde \tau_L}e^{-\Gamma_R\tilde \tau_R}},
\eeq
and an analogous expression for $S_L(\tilde\tau_L)$.

The amount of heat pumped out of the cold ($L$) reservoir during each
cycle is calculated by substituting the derivative of $S_L(t)$ (Eq.~(\ref{eq:SK}))
into Eq.~(\ref{eq:QWSymm}), then integrating over the contact time
with this reservoir,
\bea \mathcal{Q}_L&=& \omega_L \left(
S_L(0)-S_L^{eq}\right)(e^{-\Gamma_L\tilde \tau_L}-1) \nonumber\\
&=& \omega_L\left(
S_L^{eq}-S_R^{eq}\right)\frac{(1-e^{-\Gamma_L\tilde
\tau_L})(1-e^{-\Gamma_R\tilde \tau_R})}{(1-e^{-\Gamma_L\tilde
\tau_L}e^{-\Gamma_R\tilde \tau_R})}.
 \label{eq:Q1} \eea
When the coupling times $\tilde \tau_L$ and $\tilde \tau_R$ are
 long relative to the inverse relaxation
rates, the TLS equilibrates with the heat baths during the thermal
branches. Then the heat pumped per cycle is maximized
\beq \mathcal{Q}_L= \omega_L \left(S_L^{eq}-S_R^{eq}\right).
\label{eq:Qopt} \eeq
Based on this equation we can derive the condition for attaining
the desired pumping action: $\mathcal Q_L$ is required to be positive, implying
that $S_L^{eq}>S_R^{eq}$.
In the classical limit, $\omega_K< k_B T_K$,  using
Eq.~(\ref{eq:Seq}), this translates into
the condition $\omega_L/\omega_R < T_L/T_R$.
The work performed on the system can be calculated similarly to yield
\cite{Feldmann}
\bea \mathcal{W} &=&(\omega_R-\omega_L)(S_L(\tilde \tau_L)-S_R(\tilde \tau_R)) \nonumber\\
&=&(\omega_R-\omega_L)(S_L^{eq}-S_R^{eq})
\frac{\left(1-e^{-\Gamma_L\tilde
\tau_L}\right)\left(1-e^{-\Gamma_R\tilde \tau_R}\right)}
{1-e^{-\Gamma_L\tilde \tau_L}e^{-\Gamma_R\tilde \tau_R}}.
\label{eq:WI}
 \eea
When $\Gamma_K \tilde \tau_K \rightarrow \infty$, the work
approaches
\beq
\mathcal{W}= (\omega_R-\omega_L)(S_L^{eq}-S_R^{eq}).
\label{eq:Wopt}
\eeq
The COP of this idealized machine, Eq. (\ref{eq:etaC}),  is then
\beq
\eta_C=\omega_L/(\omega_R-\omega_L).
\label{eq:etaopt}
\eeq
It does not depend on the
temperature, only on the minimal and maximal values of the
molecular energy gap.
Note that in the opposite $\tilde \tau_K\Gamma_K\rightarrow 0$
limit, expanding $e^{-\Gamma_K\tilde \tau_K}\sim 1-\Gamma_K\tilde
\tau_K$, leads to
\beq \mathcal{Q}_L, \mathcal{W} \propto \frac{\Gamma_L \Gamma_R
\tilde \tau_L \tilde \tau_R}{\Gamma_L \tilde \tau_L+\Gamma_R\tilde
\tau_R}. \eeq
If we further assume that the contact times are proportional to
the inverse of the energy gap modulation frequency, we conclude
that both $\mathcal{Q}_L$ and $\mathcal{W}$ scale like
$\Omega^{-1}$. In the next section we compare these results with
the performance of the realistic machine introduced in Section
III.

\section{ Results}

In the general case Eq.~(\ref{eq:Master}) has to be solved
numerically for the populations $P_1(t)$ and $P_0(t)=1-P_1(t)$,
and we use the fourth order Runge-Kutta method for this purpose.
We focus on the long time behavior of these quantities in order to
eliminate effects of the initial conditions. The heat current, the
applied work, and the machine efficiency are calculated using
Eqs.~(\ref{eq:QW})-(\ref{eq:etaC}). In order to retain the
markovian limit we choose a set of parameters that fulfills
$\omega_D^K \gg \Gamma$, $(K=L,R)$.
The adiabatic criteria is additionally preserved when $d \omega/dt
\ll F(t) \omega_D^K$.

\begin{figure}[htbp]
\vspace{0mm} \hspace{0mm}
 {\hbox{\epsfxsize=90mm \epsffile{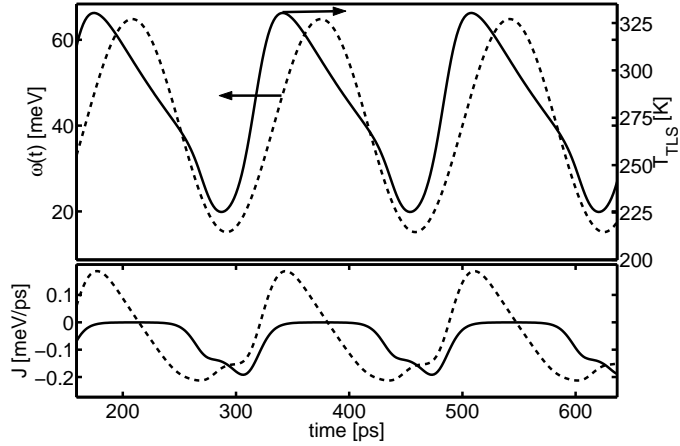}}}
 \caption{Adiabatic heat pump under a pure sine modulation, $C_1$=25 meV,
$C_{n\neq1}=0$, $A_n$=0.
 $\Omega=0.025$ meV,
$\omega_0=40$  meV,
 $\kappa_L=2.5 $ meV, $\kappa_R=0.1$  meV,
$\omega_D^L=6$  meV, $\omega_D^R=250$ meV,
 $T_L=200$ K, $T_R=300$ K.
  Top: Energy spacing (dashed line, left vertical axis) and the resulting TLS temperature calculated using
Eq.~(\ref{eq:TTLS}) (full line, right vertical axis).
  Bottom: heat current $J_L=\dot{\mathcal{Q}_{L}}$
(full) and $J_R=-\dot{\mathcal{Q}_{R}}$ (dashed).}
 \label{FigAd}
\end{figure}

We begin by analyzing an {\it adiabatic} machine operating under a
pure sine modulation of the TLS gap,
 $A_n=0$, $C_1=$25 meV, $C_{n\neq1}=0$, in Eq.~(\ref{eq:Fom}).
The choice of $\Omega$=0.025 meV, $\omega_D^L$=6 meV and $\omega_D^R$=250 meV
 corresponds to the adiabatic limit.
The  rate constants are therefore calculated using equations
(\ref{eq:rateAd})-(\ref{eq:expad}) instead of the general
expressions (\ref{eq:I1})-(\ref{eq:expnad}). This simplifies
significantly the computational effort, since for $C_1/\Omega \sim
1000$ expansion terms $J_n$ up to the order $n\sim$1200 have to be
taken into account in order to achieve convergence. We have also
verified that the adiabatic results perfectly agree with the
general formalism. The results of this calculation for this
choice of parameters (other parameters are noted at the caption)
are displays in Fig. \ref{FigAd}.
Shown are the TLS spacing modulation, the instantaneous TLS
temperature and the instantaneous heat transferred at the cold and
hot interfaces.
We find that in this adiabatic limit the device does not pump heat,
and energy is transferred from {\it both} the hot bath and the
external periodic field into the cold reservoir. This is also
demonstrated through the temperature of the TLS which is always
higher than the temperature of the cold bath. Therefore,
extraction of heat from the $L$ reservoir is impossible, and $J_L$
is negative throughout the cycle.


\begin{figure}
\vspace{0mm} \hspace{0mm}
{\hbox{\epsfxsize=90mm \epsffile{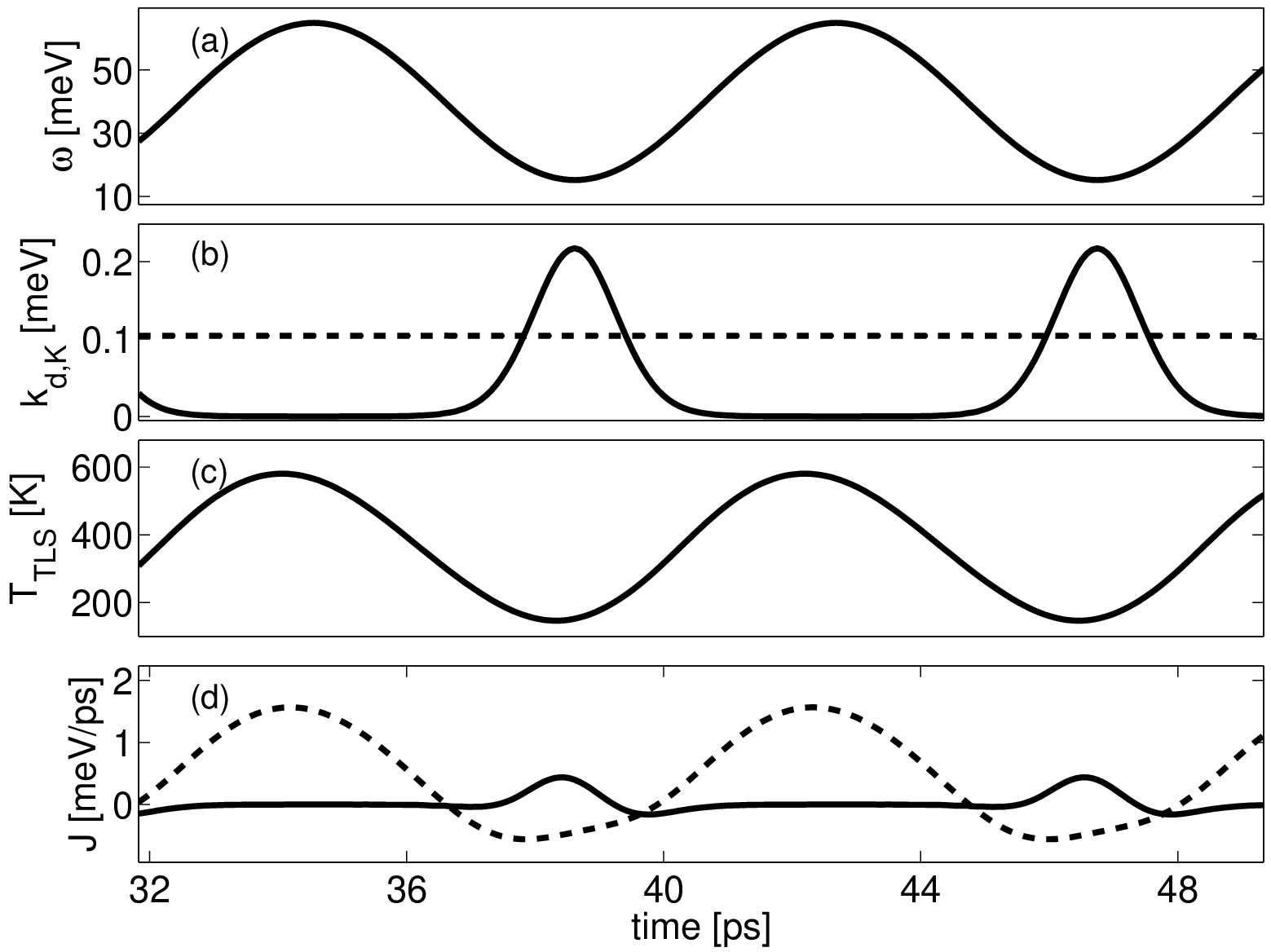}}}
\caption{ Quasi adiabatic heat pump operating under the modulation
frequency $\Omega=0.5 $ meV. Other parameters are as in Fig.
\ref{FigAd}. Shown are the TLS energy spacing modulation (a), the
relaxation rates at the right (dashed) and left (full) contacts
(b),  the TLS temperature (c),
 the heat currents $J_L$
(full) and $J_R$ (dashed) (d).}
\label{FignAd}
\end{figure}

Consider next the {\it quasiadiabatic} situation where the TLS energy
spacing is modulated at a frequency that is at the same order or smaller than
the inverse of the reservoirs relaxation times.
In Fig.~\ref{FignAd} we display the behavior
of such a machine, where the parameters are the same as in the
previous adiabatic model except that
 $\Omega$=0.5 meV.
 In panels (a)-(d) we show
(a) The (periodic) time variation of the TLS energy spacing
$\omega(t)$ .
 (b) The effective couplings $k_{d,L}$ and $k_{d,R}$
of the TLS to the left and right reservoirs.
 When $\omega(t)$ reaches its maximum
value, $k_{d,L}$ becomes negligible, $\sim$ 1$\times$10$^{-4}$
meV, and the TLS is effectively disconnected from the $L$
reservoir. In contrast, since the Debye frequency at the $R$ side
is significantly larger than the molecular frequencies, $k_{d,R}$
remains effectively constant at all times. (c) The TLS
temperature. When $\omega(t)$ becomes large, the TLS temperature reaches
a maximum of $\sim$ 600 K, larger than $T_R$=300 K.
Heat transfers then from
the hot molecular mode to the right reservoir. The lowest TLS
temperature of $\sim 150$ K is obtained at small energy spacing,
$\omega$=15 meV, at which $k_{d,L}\sim$ 0.2 meV and
$k_{d,R}=$ 0.1 meV. Therefore at this point the molecular mode
gets heat from both $L$ and $R$ reservoirs. This is seen in the
bottom panel (d):  $J_{L}$  (full line) is positive and  $J_{R}$
(dashed line) is negative
when the energy gap $\omega$ is in the neighborhood of this small value.

For this operation mode and for these engine parameters we find
that the amount of heat pumped out of the cold reservoir at each
cycle is $\mathcal Q$=0.23 meV, and the efficiency of the machine is
$\eta_{C}$=0.073. Note that the heat pumped into the right contact
{\it is not} the same
due to the action of the external force. In addition,
looking at the instantaneous pumping
we observe a delay of half a cycle in the pumping
action: Heat is pumped from the left cold reservoir when the TLS
gap is minimal, and it is injected  into the right hot reservoir
after half of a cycle when $\omega(t)$ becomes large. In between,
due to the slow decoupling rate of the $L$ reservoir from the
molecule, a backward heat flow is observed.

\begin{figure}
\vspace{0mm} \hspace{0mm} {\hbox{\epsfxsize=80mm
\epsffile{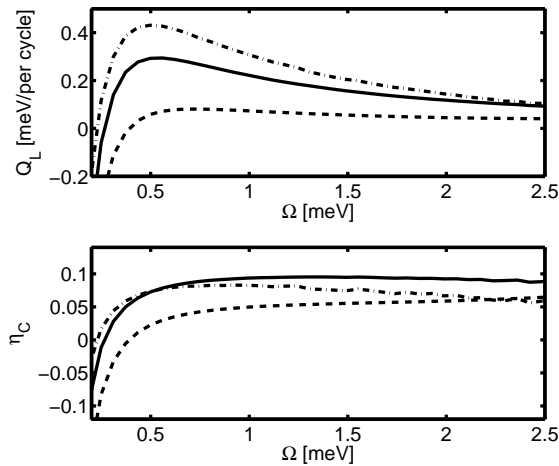}}}
\caption{Efficiency of the non adiabatic heat pump displayed in
Fig.~\ref{FignAd}. Top: Heat transferred per cycle out of the cold
reservoir. Bottom: The refrigerator coefficient of
performance (Eq.~(\ref{eq:etaC})). $C_1=$20
meV (dashed), $C_1=$25 meV (Full), $C_1=$30
meV (dashed-dotted). } \label{FigE1}
\end{figure}

Figure 4 shows results pertaining to the efficiency of this
machine. The top panel of Fig.~\ref{FigE1} presents the heat
transferred per cycle, plotted against the driving frequency. It
reaches a maximal value for $\Omega\sim $0.5 meV. The efficiency
defined in Eq.~(\ref{eq:etaC}) increases monotonically, saturates,
then decays very slowly. We can explain these observations as
follows: For very slow modulation  the TLS is at steady state
driven by its coupling to the  $L$ and $R$ reservoirs and its
temperature is approximately given by $T_{TLS}\sim (\Gamma_L T_L
+\Gamma_R T_R)/(\Gamma_L+\Gamma_R)$ \cite{Spinboson}, higher than
$T_L=T_C$. Heat then always flows towards the cold bath and
$\mathcal{Q}_L$ is negative. The finite coupling to both
reservoirs at all times therefore {\it inhibits} the pumping
operation in the adiabatic regime. This is in fact the extreme
opposite to the optimal situation in which the system is decoupled
from the reservoirs whenever needed, that leads to Eq.
(\ref{eq:Qopt}).
In the opposite {\it fast} modulation limit the TLS temperature can
reach values below $T_C$, and can pump heat out of the cold
reservoir as in Eq. (\ref{eq:Q1}). Its efficiency is however
restricted by the fact that time is insufficient for a full
equilibration with the cold reservoir, thus the total energy
injection is small.
We have also verified that in this regime
both heat and work decay like $\Omega^{-1}$. This implies that
maximal heat pumping is obtained at some intermediate modulation
frequency, as seen in the upper panel of Fig. \ref{FigE1}.
The machine performance is therefore optimal when working in the
{\it quasiadiabatic} regime.


\begin{figure}
\vspace{0mm} \hspace{0mm} {\hbox{\epsfxsize=80mm
\epsffile{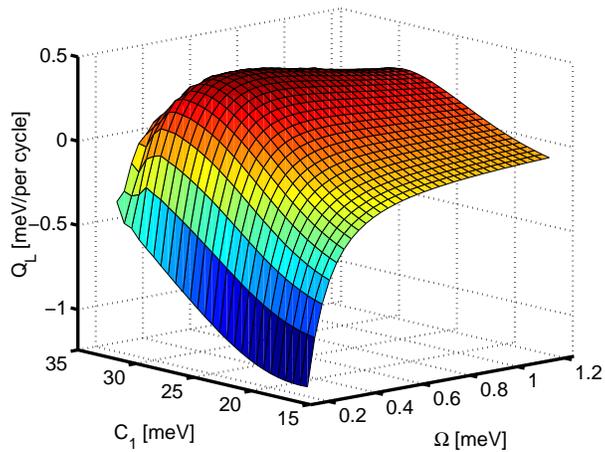}}}
\caption{ (Color online) A surface plot of the heat pumped displayed against the driving
frequency $\Omega$ and amplitude $C_1$ for a system characterized
by the parameters of Fig.~\ref{FignAd}. }
\label{FigE2}
\end{figure}

\begin{figure}
\vspace{0mm} \hspace{0mm} {\hbox{\epsfxsize=90mm
\epsffile{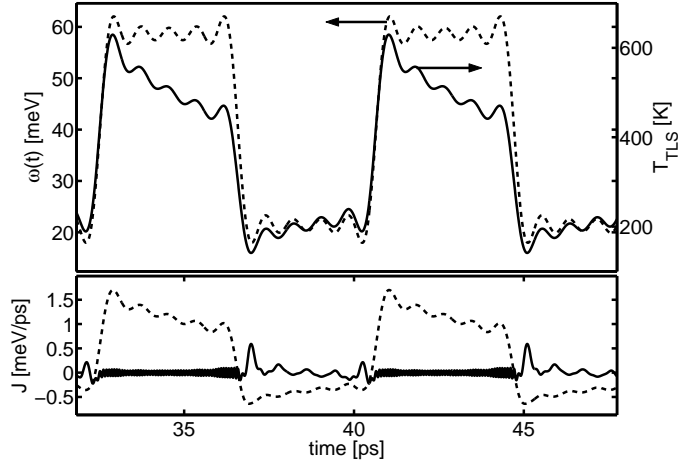}}}
\caption{ Non-adiabatic heat pump. $C_n=4/{\pi n}\times 18$ meV, $n$=1..9,
$A_n$=0, $\Omega=0.5$ meV,
$\omega_0=40$  meV,
 $\kappa_L$=2.5 meV, $\kappa_R=0.1$  meV,
$\omega_D^L=6$  meV, $\omega_D^R=250$ meV,
 $T_L=200$ K, $T_R=300$ K.
 (Top) Shown are the TLS energy spacing modulation (left)
and the resulting TLS temperature (right).
The bottom panel displays the heat currents $\dot{\mathcal{Q}_{L}}$
(full) and $-\dot{\mathcal{Q}_{R}}$ (dashed).}
\label{FigOp}
\end{figure}

The results obtained above make it possible to investigate ways
to optimize the performance of our heat pump.  Figure \ref{FigE2}
presents a surface plot of $\mathcal{Q}_L$ per cycle as a function
of amplitude $C_1$ and frequency $\Omega$ for a sine type
perturbation. We find that for very weak modulations the system
cannot pump heat, i.e. $\mathcal{Q}_L$ is negative. The maximal
amount of heat per cycle of $\mathcal {Q}_L$=0.4 meV is
pumped at $C_1\sim 30$ meV and $\Omega \sim 0.5$ meV.

Another technique for optimizing the heat pump operation is by
devising an optimized shape for the modulation function $F(t)$. As
discussed above, $F(t)$ should be designed so as to minimize
 reverse heat transport processes. The parameters that
can be manipulated are the functional form of the modulation, the
total time duration of the pulse, and the time allocated to the
four operation branches \cite{Feldmann}.

As displayed in  Fig.~\ref{FignAd}, when utilizing a sine
modulation with the given amplitude and coupling parameters, the
TLS temperature varies between the minimal value of $T$=150 K
$<T_L$ and the maximal value $T$=600 K $> T_R$. We have argued
that the machine efficiency can be improved if  the
thermal branches are long enough so that the TLS comes as close as
possible to equilibrium  with the corresponding thermal bath,
and if  between the thermal
branches the energy gap is varied as rapidly as possible.
In Figure \ref{FigOp}
we show such a machine where the modulation is tailored such as to
"wait" at its maximal and minimal values giving the TLS
more time to equilibrate with the different reservoirs at
different parts of its cycle.
At the same time the energy gap is changed rapidly in order to
eliminate backward flow to the $L$ bath. The top panel (dashed
line) presents the shape of the driving signal and the ensuing TLS
temperature. This modulation function is constructed from the
series $F(t)=\sum_{n=1,3..L_n}C_n\sin(\Omega n t)$, $C_n=4/\pi n
\times 18$ meV, $L_n$=9. We find that indeed during the
"wait" time at the minimal energy spacing $\sim \omega=20$ meV
the TLS heats from 150 K up to $\sim$200 K. The same effect is
observed when the TLS reaches its maximal $\omega=60$ meV
value: The TLS temperature reduces from $\sim 600$ K down to
$\sim$ 400 K. The  fast oscillations of $J_L$ at high temperatures
(e.g. between 40-45 ps) sum up close to zero. The refrigerator
efficiency is $\eta_C$=0.021 and $\mathcal{Q}_L$=0.06 meV
per cycle. The corresponding machine operating under a pure sine
modulation with the same amplitude ($L_n$=1, $C_1$=18 meV)
does not pump heat.


\section{Summary}
In analogy with electron and spin pumps that were
investigated in recent years, we propose here a molecular level
thermal machine that pumps energy from a cold to a hot reservoir.
We discuss its operating principles, performance, optimization
techniques, and possible physical realizations. We expect that in
the future devices whose functionality is determined by both their
thermal and electrical properties \cite{Humphrey} will be of great
interest.


\renewcommand{\theequation}{A\arabic{equation}}
\setcounter{equation}{0}  
\section*{APPENDIX A: Derivation of the quantum master equation.}

Here we derive the quantum master equation for a two level system
with a time dependent energy spacing which is interacting with two
thermal reservoirs, Eqs.~(\ref{eq:Hamilt})-(\ref{eq:HMB})
\bea
 H&=&\frac{1}{2}\left( \omega_0 + F(t) \right)
\left( |1\rangle \langle 1| - |0 \rangle \langle 0| \right)
\nonumber\\
&+&B|0\rangle \langle 1| + B^{\dagger}|1\rangle \langle 0| +H_B.
\eea
$H_B=H_L+H_R$, and the system bath coupling $B$ includes the
$L$ and $R$ terms, $B=B_L+B_R$. The evolution of the total density matrix
is given by the Liouville equation ($\hbar\equiv1$)
\beq
\frac{\partial\rho}{\partial t}=-i[H,\rho].
\eeq
The equations of motion for each density matrix component $\rho_{i,j}$ are
given in terms of the bath operators
\bea
\dot \rho _{1,1}&=&-iB^{\dagger}\rho_{0,1} +i\rho_{1,0}B,
\nonumber\\
\dot \rho _{0,0}&=&-iB\rho_{1,0} +i\rho_{0,1}B^{\dagger},
\nonumber\\
\dot \rho _{0,1}&=&i\left(\omega_0+F(t) \right)
\rho_{0,1}-iB\rho_{1,1} +i\rho_{0,0}B,
\nonumber\\
\dot \rho _{1,0}&=&-i\left( \omega_0+F(t)\right)\rho_{1,0} -iB^{\dagger}\rho_{0,0} +i\rho_{1,1}B^{\dagger}.
\label{eq:Li}
\eea
Next we formally integrate the nondiagonal terms $\dot\rho_{0,1}$ and
$\dot\rho_{1,0}$ using the Leibnitz integral rule
\bea \frac{d}{dt}\int_{u(t)}^{v(t)} f(t,\tau)d\tau=
 v'(t)f(t,v(t))-u'(t)f(t,u(t))
\nonumber\\
+\int_{u(t)}^{v(t)}\frac{\partial}{\partial t}f(t,\tau)d\tau,
\eea
and obtain
\bea
\rho_{0,1}(t)&=&\int_{0}^{t}e^{i\omega_0(t-\tau)} e^{i[f(t)-f(\tau)]}
\left[ -iB(\tau) \rho_{1,1}(\tau) +i\rho_{0,0}(\tau)B(\tau)\right]d\tau.
\nonumber\\
\rho_{1,0}(t)&=& \rho_{0,1}^*(t)
\eea
where $f(t)=\int F(t) dt$.
We substitute these expressions into the equations of the diagonal terms
$\dot\rho_{0,0}$ and $\dot\rho_{1,1}$ and trace over both
$L$ and $R$ thermal baths
assuming the density matrix can be decomposed at all times by
$\rho(t)=\rho_{L}\rho_{R} \sigma(t)$.
 Here $\sigma$ is the reduced density matrix operator and
$\rho_K=e^{-\beta_K H_K}/{\rm Trace} (e^{-\beta_K H_K})$, $K=L,R$.
Following the standard Redfield-Bloch derivation \cite{Redfield},
i.e. second order perturbation theory combined with the assumption
that bath correlation functions decay rapidly on the time scale of
the change of $\sigma$, we obtain the quantum master
equation for the diagonal reduced density matrix elements
$P_n=\sigma_{n,n}$, $n$=0,1
\bea \dot P_{1}&=&P_{0}(t) \int_{0}^{t}
e^{i\omega_0(t-\tau)} e^{i[f(t)-f(\tau)]}
\langle B(\tau)B^{\dagger}(t)\rangle d\tau
\nonumber\\
&+& P_{0}(t) \int_{0}^{t} e^{-i\omega_0(t-\tau)}
e^{-i[f(t)-f(\tau)]}
\langle B(t)B^{\dagger}(\tau)\rangle d\tau
\nonumber\\
&-& P_{1}(t) \int_{0}^{t} e^{i\omega_0(t-\tau)}
e^{i[ f(t)-f(\tau)]}
\langle B^{\dagger}(t)B(\tau) \rangle d\tau
\nonumber\\
&-&
 P_{1}(t) \int_{0}^{t} e^{-i\omega_0(t-\tau)}
e^{-i[f(t)-f(\tau)]}  \langle
B^{\dagger}(\tau)B(t)\rangle d\tau;
\nonumber\\
 P_{0}(t)&=&1- P_{1}(t).
\label{eq:rate}
\eea
In the markovian limit we further extend the upper limit in the
integrals to infinity, and assume that bath correlation functions
do not depend on the initial time. Note that no restrictions are
imposed on the modulation term $F(t)$, e.g. in general it need not
be periodic.


\end{document}